\documentclass[]{article}  
\usepackage[]{graphicx}
\usepackage{setspace}
\usepackage[]{graphicx}
\usepackage{amsmath}
\usepackage{amssymb}
\usepackage{subfigure}
\usepackage{epsfig}
\usepackage{wrapfig}
\usepackage{color}
\usepackage{rotating}
\usepackage{float}

\def\##1{{\bf #1}}
\def\=#1{\underline{\underline #1}}

\def\.{\mbox{ \tiny{$^\bullet$} }}

\def\lambdao{\lambda_{\scriptscriptstyle 0}}

\begin{document} 
\begin{center}
\noindent{\bf Experimental investigation of circular Bragg
phenomenon exhibited  by a mirror-backed  chiral sculptured thin film} \\[6pt]
 
\noindent{\it Sema Erten, Stephen E. Swiontek, Christian M. Graham and \\ Akhlesh Lakhtakia{$^\ast$}}\\[6pt]

\noindent {Pennsylvania State University, Department of Engineering Science and Mechanics, 212 EES Building,  University Park, PA 16802, USA\\
\noindent$^\ast$ {akhlesh@psu.edu} }
\end{center}

 \begin{abstract}
Experimentation with obliquely incident light
established that all four circular reflectances of a chiral sculptured thin film backed by a metallic mirror contain strong evidence
of the circular Bragg phenomenon. 
When the mirror is removed, strong evidence of that phenomenon  is found only in the spectrum of 
the co-polarized and co-handed reflectance. \end{abstract}

\section{Introduction}

The permittivity dyadic of a periodic structurally chiral material, exemplified by cholesteric liquid crystals \cite{Fergason,Chandra,deG} and chiral sculptured thin films (STFs) \cite{STFbook}, rotates at a fixed rate about a fixed axis. The consequent periodic unidirectional nonhomogeneity is responsible for the exhibition of the circular Bragg phenomenon \cite{FLaop} by the structurally chiral material. This phenomenon can be described as follows: When a circularly
polarized plane wave is  incident on a periodic structurally chiral material
of sufficient thickness $D$ {embedded in free space}, the direction of nonhomogeneity is the same as
the thickness direction, the angle of incidence $\theta$ with respect to
the thickness direction is not very high, and the free-space wavelength $\lambdao$ lies
in a specific regime called the circular Bragg regime, the reflectance is 
\begin{itemize}
\item 
high if the handedness of the incident plane
wave is identical to that of the material, {but
\item
low if the two handednesses are different.}
\end{itemize}
If the periodic structurally chiral material is weakly dissipative, low/high reflectance {in}  the circular Bragg regime is accompanied by high/low transmittance. Particularly for normal incidence (i.e., $\theta=0$), the  circular Bragg
phenomenon is  exploited for circular-polarization filters, spectroscopy, and optical-sensing
applications  \cite{FLaop,Jacobs,STFbook}.
 
 Would the circular-polarization selectivity in the circular Bragg regime exist even if transmission were to be completely thwarted by backing the periodic structurally chiral material  with a  highly reflecting standard mirror such as a highly polished metal sheet? To our knowledge, no report on this question has been published.  

In this Letter, we report the measured circular reflectances of a chiral STF made of zinc selenide and backed by a silver mirror. The angle of incidence $\theta$ was varied from $10$~deg  to $70$~deg while the wavelength $\lambdao$  was varied between $600$~nm and $900$~nm. Both ranges were sufficient to comprehensively accommodate the circular Bragg phenomenon for the chosen material.
 
\section{Materials and methods}
 A structurally right-handed chiral STF was deposited on a pre-cleaned microscope glass slide (82027-788,
VWR, Radnor, PA, USA) in a
 low-pressure chamber (Torr International, New Windsor, NY, USA) as follows.
Two grams of powdered zinc selenide (Alfa Aesar, Ward Hill, MA, USA)
were loaded into a tungsten boat (S6-.005W, R. D. Mathis, Long Beach, CA, USA) which was secured as the heating element
in an electrical heater located inside the low-pressure chamber.  The glass slide and a silicon wafer were affixed
side by side to a rotatable planar platform
about 15~cm above the boat. A quartz crystal monitor   was mounted close to the
platform to help control the deposition rate. The platform was oriented 
 so that  a collimated vapor flux of zinc selenide would be directed during deposition at  $20$~deg  with respect to
the plane of the platform.
The chamber was closed and pumped down to a base pressure of $6.5$~$\mu$Torr. Then a $95$-A current was
passed through the tungsten boat.  The serial bideposition technique 
\cite{HWKLR} was implemented
while maintaining the deposition rate at $0.4\pm0.02$~nm~s$^{-1}$.  A 6-{periods}-thick chiral STF was grown. The procedure was repeated  without removing the glass slide and the silicon wafer but after replenishing the boat with zinc selenide so that the chiral STF 
{finally had 12 periods along the thickness direction.}

\begin{figure}[htbp]
\begin{center}
\includegraphics[scale = 0.25]{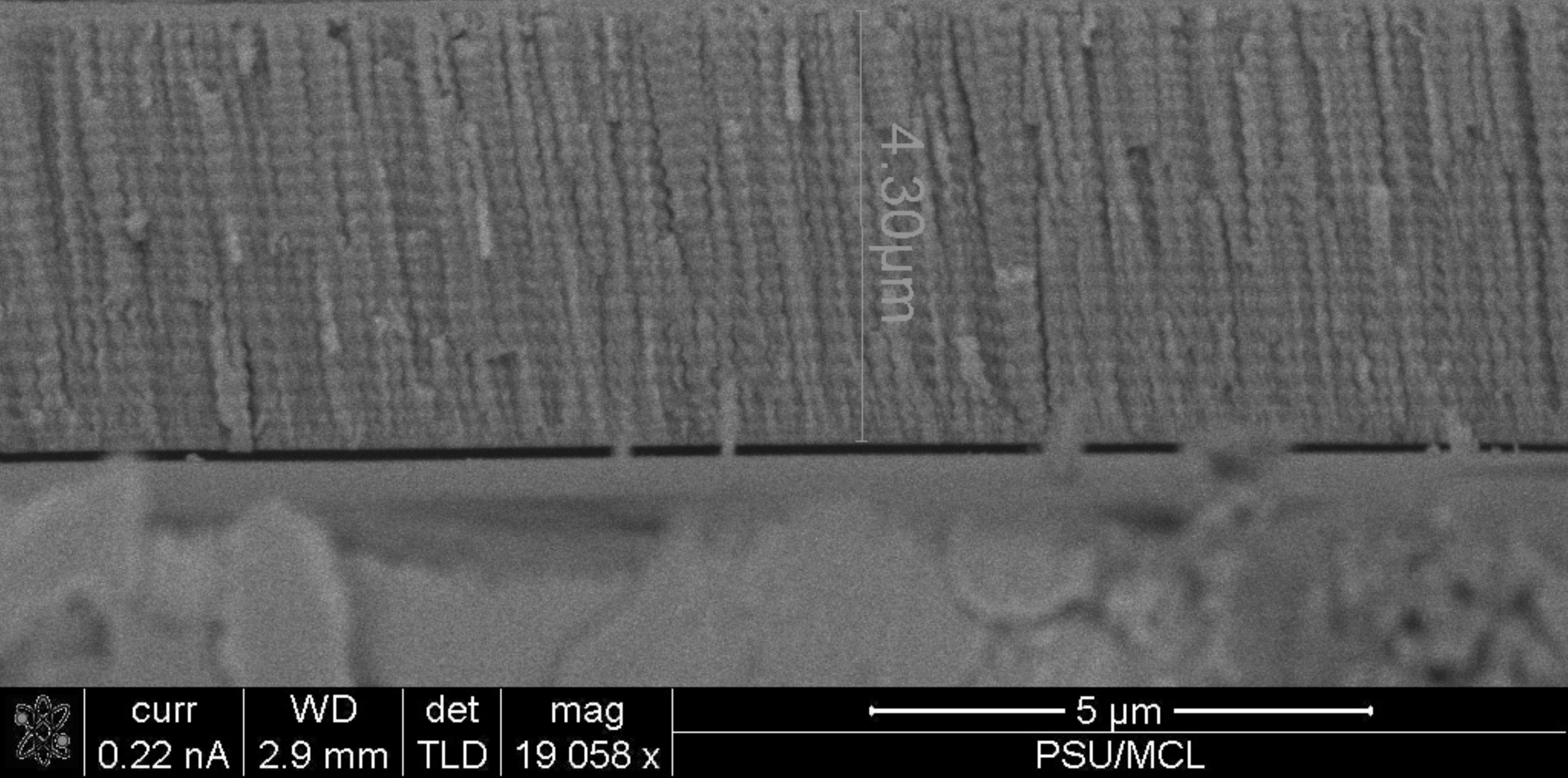}
\caption{Cross-sectional image on a scanning electron microscope of the  12-{periods}-thick  chiral STF fabricated on
a silicon wafer.
}
\label{SEMimage}
\end{center}
\end{figure}

Cross-sectional images of the  12-{periods}-thick  chiral STF grown on the silicon wafer were acquired on a scanning electron microscope (FEI Nova NanoSEM 630, Hillsboro, OR, USA). Structural chirality is
clearly evident in the representative image shown in Fig.~\ref{SEMimage}.   As the  thickness  $D=4.30$~$\mu$m, the period of the chiral STF is $D/12=358.33$~nm.

A custom computer-controlled  variable-angle  spectroscopic system  was used to measure  
the circular reflectances ($R_{RR}$, $R_{LL}$, $R_{RL}$, and $R_{LR}$)  of the chiral STF on the glass slide
as functions of $\theta$ and $\lambdao$. The first subscript in $R_{LR}$ 
indicates the \textit{L}eft-circular polarization (LCP) state of the reflected light and the second subscript indicates the
\textit{R}ight-circular polarization (RCP) state of the incident light, and similarly for the other three circular reflectances. {Co-polarized reflectances have both subscripts identical, but cross-polarized reflectances do not.}

Light from
a halogen light source (HL-2000, Ocean Optics, Dunedin, FL, USA) was
passed first through a Glan--Taylor linear polarizer (GT10, ThorLabs, Newton, NJ, USA) 
positioned inside a lens tube mount  (SM1PM10, ThorLabs) inserted in a rotation mount (PRM1Z8, ThorLabs)
and then through a Fresnel rhomb (LMR1, ThorLabs), before impinging on the chiral STF. The polarizer was set  so that the light incident on the chiral STF possessed  either the RCP or the LCP state. The chiral STF was mounted on a rotatable stage to change the angle $\theta\in[{10}\deg,70\deg]$ as desired. The light specularly reflected by the chiral STF traveled
through a Fresnel rhomb and a Glan--Taylor linear polarizer to a CCD spectrometer (HRS-BD1-025, Mightex Systems, Pleasanton, CA) to measure the intensity. The second polarizer was set  so that the light incident on the CCD spectrometer possessed  either the RCP or the LCP state. The reflected intensity was divided by the previously measured incident intensity to obtain the circular reflectance. We measured all four circular reflectances, first of the chiral STF 
 on the glass slide alone and then of the chiral STF  on the glass slide
backed by a silver mirror (PFSQ10-03-P01, Thorlabs).

 \section{Results and discussion}\label{nrd}

\begin{figure}[htbp]
\begin{center}
\includegraphics[scale = 0.4]{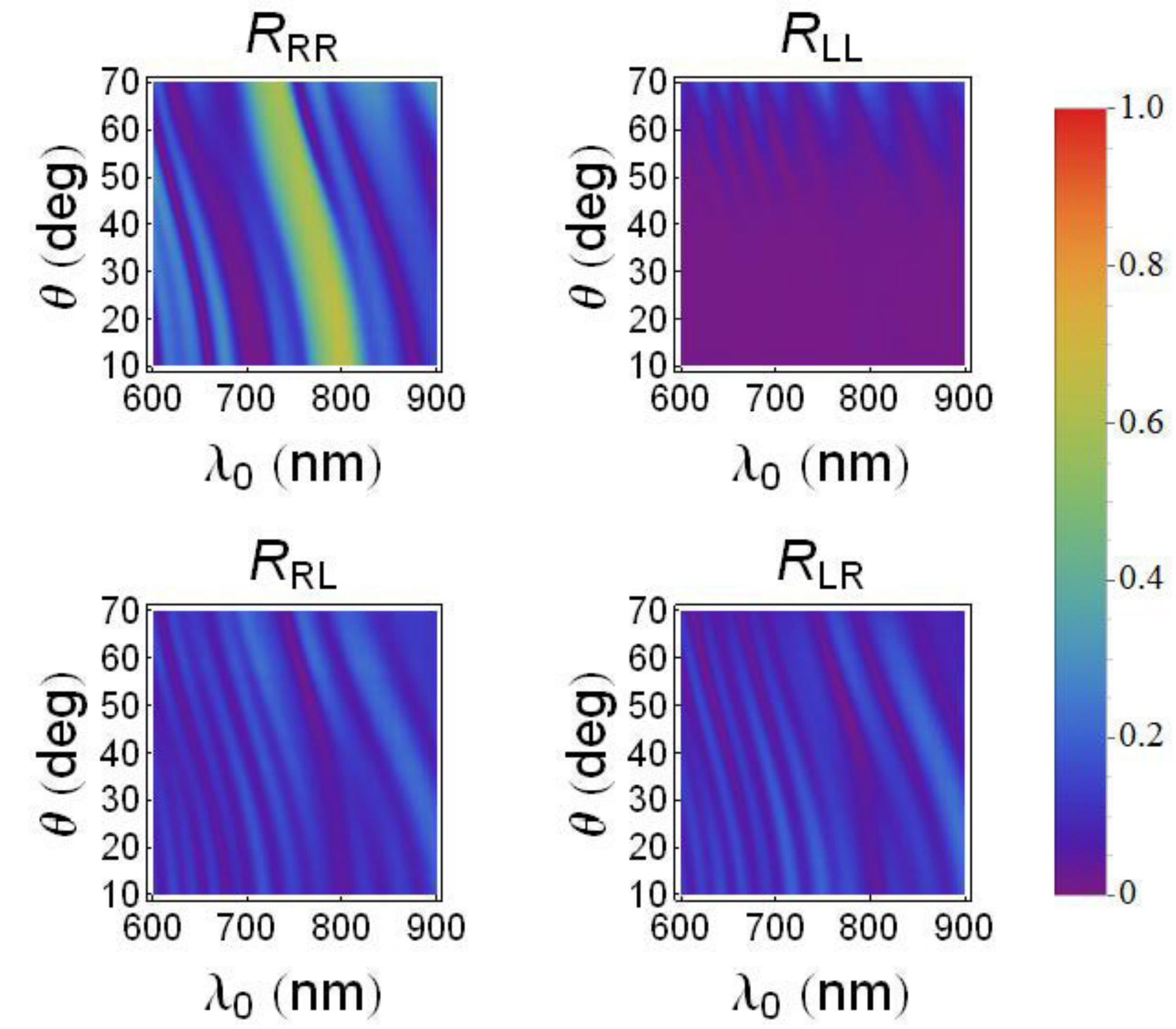}
\caption{Measured spectrums for   $\theta\in[10~{\rm deg},70~{\rm deg}]$ of the four circular reflectances of the {12-periods}-thick chiral STF fabricated on glass.}
\label{Fig2}
\end{center}
\end{figure}

Measured spectrums of the four circular reflectances of the
{structurally right-handed} chiral STF deposited on the glass slide are
shown in Fig.~\ref{Fig2} for    $\theta\in[10~{\rm deg},70~{\rm deg}]$.
{The circular Bragg regime
 is clearly manifested in the spectrum of $R_{RR}$ as a high-intensity ridge
that blue shifts as $\theta$ increases, no similar feature being evident
in the spectrum of $R_{LL}$ \cite{STFbook,FLaop,Takezoe,StJ}.} Indeed,  $R_{RR}>>R_{LL}$ in this regime,
which has
 center wavelength $796.1$~nm and
full-width-at-half-maximum   bandwidth $53.1$~nm when $\theta=10\deg$.
The center wavelength blue shifts to $721.8$~nm as $\theta$ increases to $70\deg$,
as determined from the experimental  $R_{RR}$-data using the scheme described elsewhere \cite{ELB2015}.
Even though the cross-polarized 
circular reflectances $R_{RL}$ and $R_{LR}$ are very small,  low-intensity troughs
identifying the circular Bragg regime are weakly evident in the spectrums of these two reflectances.
Although the reflectances could not be measured for normal incidence,
conclusions on the center wavelength and the bandwidth
of the circular Bragg regime drawn from $\theta=10$~deg can be easily extended to $\theta=0$ \cite{ELB2015}.

\begin{figure}[htbp]
\begin{center}
\includegraphics[scale = 0.4]{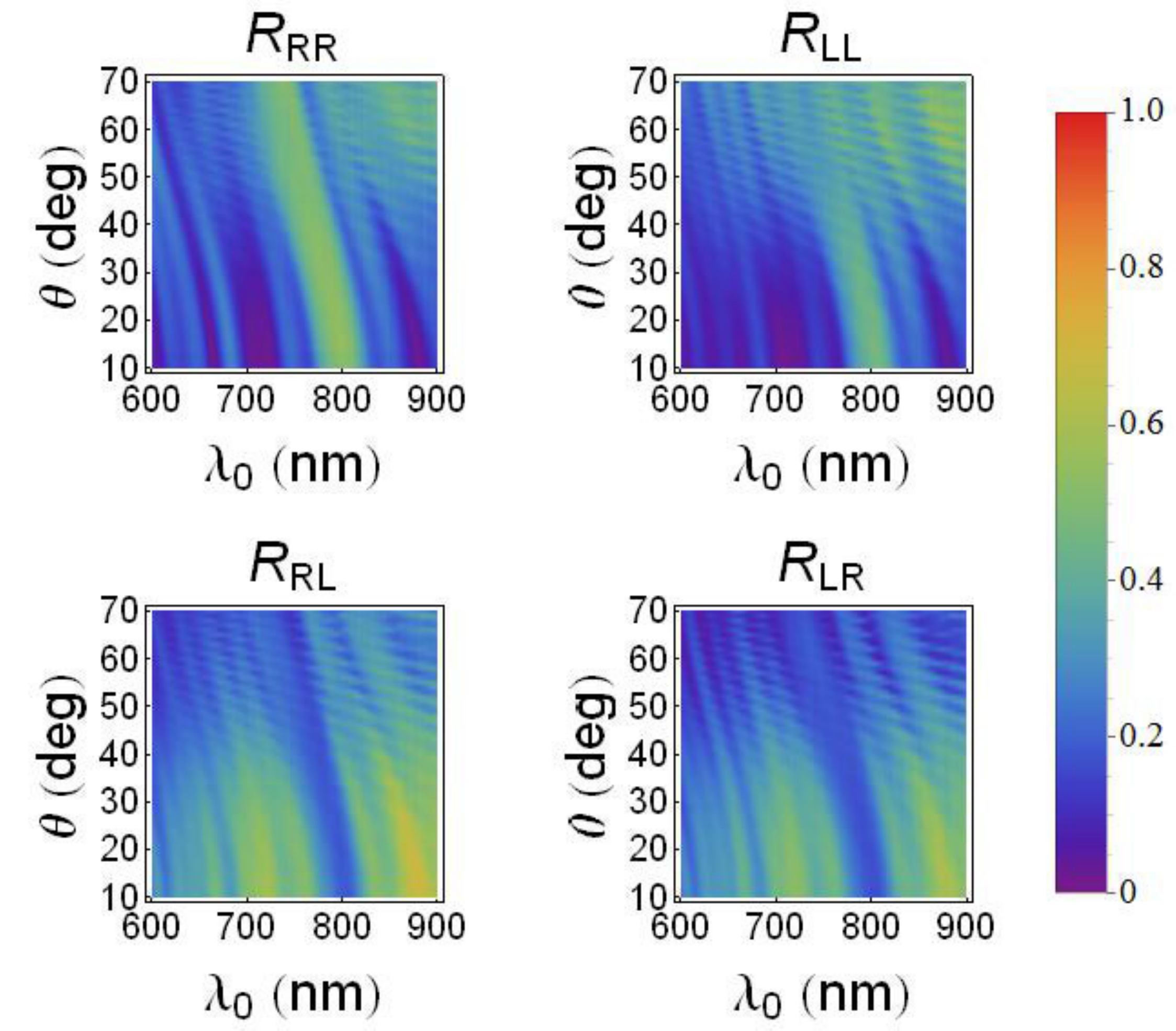}
\caption{Same as Fig.~\ref{Fig2}, but the chiral STF on glass was backed by a silver mirror.
}
\label{Fig3}
\end{center}
\end{figure}

Measured spectrums of the four circular reflectances of the
zinc-selenide chiral STF backed by a silver mirror are
shown in Fig.~\ref{Fig3} for    $\theta\in[10~{\rm deg},70~{\rm deg}]$.
The circular Bragg regime is evident in the spectrum
of $R_{RR}$ as a high-intensity ridge that blue shifts
with increasing $\theta$---similarly to the high-intensity ridge in the spectrum of $R_{RR}$ in Fig.~\ref{Fig2}. 
The center wavelength blue  shifts from $799.8$~nm to $726.6$~nm as $\theta$ increases
from ${10}\deg$ to $70\deg$, also as determined from the experimental  $R_{RR}$-data  \cite{ELB2015}.

The circular Bragg regime is  evident in Fig.~\ref{Fig3} also as a high-intensity ridge
in the spectrum of $R_{LL}$. Although $R_{RR}$ still exceeds $R_{LL}$, the difference
$R_{RR}-R_{LL}$ is considerably smaller than in Fig.~\ref{Fig2}. Furthermore, 
low-intensity troughs in the spectrums of $R_{RL}$ and $R_{LR}$ in Fig.~\ref{Fig3}
are additional footprints of the circular Bragg phenomenon, these troughs being far more prominent
than in Fig.~\ref{Fig2}.

\begin{figure}[htbp]
\begin{center}
\includegraphics[scale = 0.6]{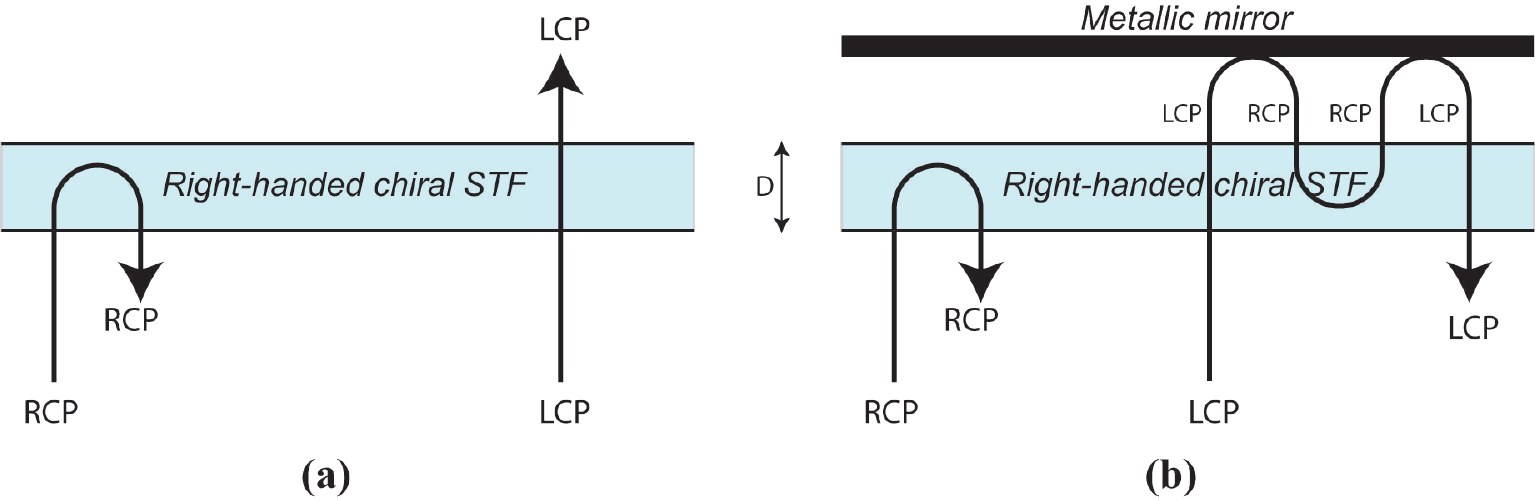}
\caption{Cartoons to explain the circular Bragg phenomenon exhibited {by} a right-handed
chiral STF (a) without and (b) with a perfect mirror behind, {when the incidence
is normal.} Cross-polarized remittances
and co-polarized remittances of small magnitude have been ignored in these cartoons.}
\label{Fig4}
\end{center}
\end{figure}

The differences between the spectrums shown in Figs.~\ref{Fig2} and \ref{Fig3} can be explained as follows.
 Coupled-mode theory shows
that the structurally right-handed chiral STF functions as a Bragg reflector when normally incident light has
the  RCP state \cite{Belyakov,McCall}; furthermore,  time-domain
simulations indicate that the reflection takes place through a large region of the chiral
STF close to its entry pupil \cite{GLoc}, which is the reason for $D$ to be sufficiently large in comparison
to the period of an SCM for a significant manifestation
of the circular Bragg phenomenon \cite{STFbook}. For   incident  light of the LCP state, the 
right-handed chiral STF functions simply as a homogeneous material, so that  reflection is low and 
transmission is high
(if dissipation is weak). Hence, $R_{RR}>>R_{LL}$. Cross-polarized reflectances and transmittances are usually small, and
can be further decreased by the use of impedance-matching layers on both faces of the
chiral STF. The simple cartoon in Fig.~\ref{Fig4}(a)
depicts the chief features of the circular Bragg phenomenon, after ignoring the cross-polarized remittances
and with the assumption that $D$ is sufficiently large.
As $\theta$ increases, a blue shift of the circular Bragg regime occurs \cite{ELB2015},
but the simple cartoon in Fig.~\ref{Fig4}(a) remains valid to a great extent.
The foregoing characteristics are evident in the spectrums presented in Fig.~\ref{Fig2}.
The circular-polarization selectivity vanishes
at very high values of $\theta$ \cite{STFbook}, but those values of $\theta$ could not be covered in our experimental
setup.

 Over the spectral regime $600$~nm~$<\lambdao<900$~nm
and for unpolarized light, the silver mirror used by us has a reflectance of at least $0.97$ for $\theta=12\deg$
and $0.94$ for $\theta=45\deg$, according to the manufacturer. Therefore, in order to explain
the characteristics of the circular Bragg phenomenon in Fig.~\ref{Fig3}, it is safe to assume that the silver mirror
is a perfect reflector. Furthermore, the silver mirror reverses the handedness of the reflected light in relation to that of the incident circularly polarized light.

A small air gap exists between the back face of the glass slide and the silver mirror. As shown
in  Fig.~\ref{Fig4}(b),
RCP light incident normally on the mirror-backed structurally right-handed
chiral STF would be substantially reflected from within
the chiral STF, provided that $D$ is sufficiently large and the wavelength lies in the circular Bragg regime;
very little light would be incident on and be reflected by the mirror.
A substantial fraction of LCP light incident normally on the same structure would reach the air gap, be incident
on the silver mirror, and would undergo three reflections before traversing the chiral STF as RCP light---as shown
also in Fig.~\ref{Fig4}(b). Since incident RCP light would travel through the chiral STF for a shorter distance than incident LCP light would,
the latter would be absorbed more than the former, leading to $R_{RR}$ exceeding {$R_{LL}$,} but the difference between the two {co-polarized} reflectances would not be as stark as when the mirror were to be removed.

Although cross-polarized remittances and small co-polarized remittances were ignored, the foregoing simple 
explanation suffices to understand the effect of the backing mirror. The blue shift of the circular Bragg regime
with increasing $\theta$ would not be disturbed by the mirror.
Outside the circular Bragg regime, $R_{RL}$ and $R_{LR}$ would be expected to be enhanced by the reversal
of the circular polarization state by the mirror, a feature that is clearly identifiable on comparing Figs.~\ref{Fig2}
and \ref{Fig3}. That enhancement also intensifies the footprints of the circular Bragg phenomenon in
 the spectrums of $R_{RL}$ and $R_{LR}$ in Fig.~\ref{Fig3} relative to their counterparts
 in Fig.~\ref{Fig2}.

 To conclude, we have experimentally established that the {signature of the} circular Bragg phenomenon is clearly evident in the spectrums
 of all four circular reflectances of a modestly dissipative  chiral STF backed by a metallic reflector. 
When the mirror is absent, strong evidence of the circular Bragg phenomenon  exists only in the spectrum of 
the co-polarized and co-handed reflectance {(i.e., $R_{RR}$ in Fig.~\ref{Fig2})}.

\section*{Acknowledgments}
 S.E.  thanks the Turkish Ministry of National Education for financial support of her graduate studies. S.E.S. and
 A.L. are grateful to the Charles Godfrey Binder Endowment at
the Pennsylvania State University for  financial support.

\end{document}